\documentclass[useAMS,usenatbib,usegraphicx]{mn2e}

\usepackage{natbib}
\usepackage{graphicx}
\usepackage{amsmath}

\newcommand {\ie} {{i.e.}}

\newcommand {\ea} {{et~al.}}
\newcommand {\be} {\begin{equation}}
\newcommand {\ee} {\end{equation}}

\newcommand {\p} {\partial}
\newcommand {\beq} {\begin{equation}}
\newcommand {\eeq} {\end{equation}}

\title{Magnetization of jets in luminous blazars}

\author[Janiak \ea]{M.~Janiak$^{1}$\thanks{E-mail: mjaniak@camk.edu.pl}, M.~Sikora$^{1}$, and R.~Moderski$^{1}$ \\
$^{1}$Nicolaus Copernicus Astronomical Center, Bartycka 18, 00-716 Warsaw, Poland }

\begin{document} 

\maketitle

\begin{abstract}
The luminosities of many powerful blazars are strongly dominated 
by $\gamma$-rays which most likely result from Comptonization of radiation
produced outside a jet. This observation sets certain constraints on the composition and energetics of the jet as well as the surrounding quasar environment. We study the dependence of Compton dominance on jet magnetization (the magnetic-to-matter energy flux) and on the location of the 'blazar zone'. Calculations are performed  for two geometries of broad-emission-line and hot-dust regions: spherical and planar. The jet magnetization corresponding to the large observed Compton dominance is found to be $\sim 0.1 (\theta_{\rm j} \Gamma)^2$ for spherical geometries and 
$\sim 0.01 (\theta_{\rm j} \Gamma)^2$ for planar geometries, 
where $\theta_{\rm j}$ is the jet half-opening angle and $\Gamma$ is the jet Lorentz factor. This implies that jets in luminous blazars are matter dominated and that this domination is particularly strong for the flattened geometry of external radiation sources. 
\end{abstract}

\begin{keywords}
quasars: general -- radiation mechanisms: non-thermal -- acceleration of particles
\end{keywords}

\section{Introduction}
As indicated by {\it CGRO}/EGRET \citep{vMon95} and confirmed by {\it Fermi}/LAT \citep{Abd10, Ack11}, the apparent luminosities of blazars associated with flat spectrum radio quasars (FSRQ) are often  dominated by $\gamma$-rays. For most of them the ratio of the $\gamma$-ray luminosity
to synchrotron luminosity is larger than 4 and in many cases exceeds 10 \citep[see][Fig. 22]{Gio12}. A dense radiative environment in the quasar nuclei  strongly favours the external-radiation-Compton (ERC) mechanism of $\gamma$-ray production \citep[see][and refs. therein]{Sik09}. In such a case with the assumption of the 'one-zone' model the $\gamma$-to-synchrotron luminosity ratio can be approximated by the ratio of the external radiation energy density, $u_{\rm ext}'$, to the internal magnetic energy density, $u_{\rm{B}}'$, both as measured in the jet co-moving frame. 
Hence, combining  knowledge about external radiation fields, kinematics of a jet, and observationally determined Compton dominance one may estimate the intensity of the magnetic field in the blazar zone and then the flux of the magnetic energy $L_{\rm B}$. Comparison of $L_{\rm B}$ with the  total jet energy flux, $L_{\rm j} \sim L_{\gamma}/(\eta_{\rm rad}\eta_{\rm e}\eta_{\rm diss} \Gamma^2)$ can then be used to determine the sigma parameter, $\sigma$, defined to be the ratio of 
the magnetic energy flux to the matter energy flux,~\ie~$\sigma \equiv L_{\rm B}/L_{\rm kin} = (L_{\rm B}/L_{\rm j})/[1-(L_{\rm B}/L_{\rm j})]$, where $\eta_{\rm diss}$ is the fraction of $L_{\rm j}$ dissipated in the blazar zone, $\eta_{\rm e}$ is the fraction of dissipated energy channelled to accelerate electrons, and $\eta_{\rm rad}$ is the average radiative efficiency of relativistic electrons \citep{Sik13}.

Studies of the $\sigma$ parameter are important not only for better understanding of the dynamical structure and evolution of relativistic jets, but also because its value determines the dominant particle acceleration mechanism (shock vs. reconnection) and its efficiency and, therefore, should be performed in a more systematic and complete manner than so far. In particular, one should take into account such uncertainties as the location of the blazar zone and the geometry of the external radiation fields. In both cases uncertainties are still very large. The blazar zone deduced by some models is located at $\sim 300 R_{\rm g}$ \citep[see][]{ste11}, while in others it can be up to thousands of times farther \citep[see][]{Mar10}. Also the geometry of the broad-line region (BLR) and of the hot-dust region (HDR) is often considered to be spherical and not stratified, while in reality it may be very flat and significantly radially extended (for BLR see \citealp{Ves00}; \citealp{Dec08}; and \citealp{Dec11}; for HDR, e.g. \citealp{Wil13}; \citealp{Ros13}). In this paper we present results of such studies by mapping theoretical blazar spectral features as a function of distance, the geometry of external photon sources, and jet magnetization and comparing them with observations. 

Our theoretical models of broad band spectra are constructed using knowledge of the typical parameters of radio-loud quasars: BH masses, Eddington ratios, and jet powers. We assume strong coupling between protons and electrons as indicated by particle-in-cell (PIC) simulations \citep[see][]{Sir11}. Detailed model assumptions are specified and discussed in \S2. Results of our modelling of blazar spectra and their features (bolometric apparent luminosities, locations of the spectral peaks, Compton dominance, external photon sources' energy densities) dependence on distance from the BH for different $\sigma$ values and different geometries of external
radiation fields are presented in \S3. They are discussed and summarised in \S4.

\section{Model Assumptions}

\subsection{Dissipation region}

The jet is assumed to propagate  with a constant bulk Lorentz factor $\Gamma$ and to diverge conically with a half opening angle  $\theta_{\rm jet} = 1/\Gamma$. Energy dissipation and particle acceleration are assumed to take place within a distance range  $r_1-r_0 = r_0$ and proceed in the steady-state manner
\citep{Sik13}. Radiation production is followed up to distance $r_2 = 10r_1$. The emitting jet volume is divided into $s$ cells, each with the same radial size $\delta r = (r_2 - r_0)/s$. Number of cells used in calculations is not a physical model property but just a parameter responsible for numerical precision. We assumed $s=100$ which is a fair compromise between accuracy and calculation time. Assuming uniformity of matter, magnetic fields and external photon fields across the jet and within the cell thickness $\delta r$, 
the emitting jet volume is approximated as a sequence of 'point sources' (the real cell emission volume is used only to compute density of the synchrotron radiation needed to calculate the synchrotron self-Compton (SSC) luminosity). Jet radiation spectra are computed for different $r_0$ and presented as a function of the parameter $r=1.5r_{0}$.

Figure~\ref{fig:model} presents a sketch of model geometry.

\setcounter{figure}{0}
\begin{figure*}
  \includegraphics[width=.6\textwidth]{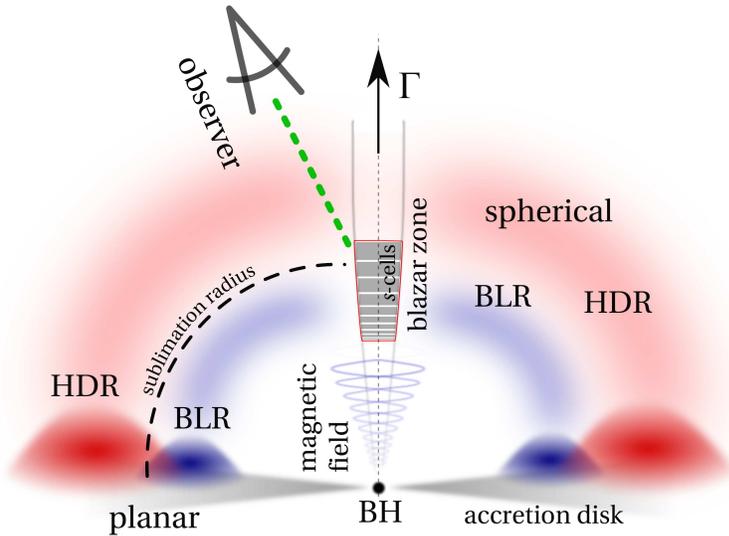}
  \caption{A sketch of blazar model geometry used in calculations.}
  \label{fig:model} 
\end{figure*}

\subsection{Electron acceleration and cooling}

We follow the evolution of electron energy distribution  
in the region of interest by solving the kinetic equation for 
relativistic electrons \citep{Mod03} which can be presented in the form
\beq
 \frac{\p N_{\gamma,i} (r)}{\p r} = - \frac{\p}{\p \gamma} \left( N_{\gamma,i}(r) \frac{{\rm d} \gamma}{{\rm d} r} \right) + \frac{ Q_{\gamma,i}(r) }{ c \beta \Gamma } \, ,
 \eeq
where $N_{\gamma,i}$ is the number of electrons per energy and cell 
volume, $\beta = \sqrt{\Gamma^2-1}/\Gamma$, 
${\rm d}\gamma/{\rm d}r= ({\rm d}\gamma/{\rm d}t')/(\beta c \Gamma)$, ${\rm d}\gamma/{\rm d}t'$ are the electron 
energy loss rates as measured in the jet co-moving frame,
and $Q_{\gamma,i}(r)$ is the electron injection function assumed to take the form
\beq 
Q_{\gamma,i}(r) = K_i f(\gamma)  \, ,
\eeq
where
\beq
f(\gamma) = \left\{ 
\begin{array}{ll}
\gamma_{\rm b}^{p_{1}-p_{2}} \gamma^{-p_{1}}\,\,\,\mathrm{for}\,\,\,\gamma_{\rm min} \leq 
\gamma \leq \gamma_{\rm b}\\
\gamma^{-p_{2}}\,\,\,\mathrm{for}\,\,\,\gamma_{\rm b} \leq \gamma \leq \gamma_{\rm max}
\end{array}
\right. .
\label{eq:deff}
\eeq
Assuming that each electron is involved in the acceleration process, one
can relate the normalisation of the injection function, $K_i$, to the
jet power using the following formula  
\beq
K_i = \left( \frac{\delta r}{r_{1}-r_{0}} \right) \left( \frac{\eta_{\rm e} 
\eta_{\rm diss} L_{{\rm j},0}} {(\bar{\gamma}_{\rm inj}-1)  m_{\rm e} c^2 
\int_{\gamma_{\rm min}}^{\gamma_{\rm max}} f(\gamma) {\rm d} \gamma} \right) \, ,
\eeq
where  
\beq
\bar{\gamma}_{\rm inj} \equiv \int_{\gamma_{\rm min}}^{\gamma_{\rm max}} \gamma f(\gamma) 
{\rm d} \gamma/\int_{\gamma_{\rm min}}^{\gamma_{\rm max}} f(\gamma) {\rm d} \gamma 
\label{eq:gammab}
\eeq 
is the average electron energy Lorentz factor and $L_{{\rm j},0}$ is the total jet power prior to the dissipation region.
Equating this to the expression for an average injected electron energy \citep{Sik13} 
\beq
\bar{\gamma}_{\rm inj} = 1+\frac{m_{\rm p}/m_{\rm e}}{n_{\rm e}/n_{\rm p}} 
\frac{\eta_{\rm e} \eta_{\rm diss}}{(1-\eta_{\rm diss})} (1-1/\Gamma)(1+\sigma) 
\label{eq:gammainj}
\eeq
gives the value of the break energy in the electron spectrum $\gamma_{\rm b}$, where $m_{\rm e}$ and $m_{\rm p}$ are the electron and proton masses, respectively, and $n_{\rm e}/n_{\rm p}$ is the pair content. For $p_1 <1$ (as expected in case of strong coupling between electrons with protons heated in the dissipation zone), $p_2 >2$ (as indicated by {\it Fermi}/LAT observations), and $\gamma_{\rm min} \ll \gamma_{\rm b} \ll \gamma_{\rm max}$, $\gamma_{\rm b}$ can be found to be of the order of $\bar{\gamma}_{\rm inj}$. 

Synchrotron, SSC and adiabatic  electron energy loss rates are calculated using the procedure presented by \citet{Mod03}, but with the energy density of synchrotron radiation given now by the formula
\beq
u_{{\rm syn},i}' = \frac {L_{{\rm syn},i}'}{2 \pi R \delta r c \Gamma} \, ,
\eeq
where $R=r \theta_{\rm j}$. The ERC electron losses are computed using the approximate formula \citep{Mod05}
\beq
\left \vert \frac{{\rm d}\gamma}{{\rm d}t'} \right \vert_{\rm ERC} =
{4 \sigma_{\rm T} \over 3 m_e c^2} \frac{ (\beta_{\rm el} \gamma)^2 u_{\rm ext}'}
{(1+b)^{3/2}} \, , \label{dotg2} 
\eeq
where $b=4 \gamma {\rm h}\nu_{\rm ext}'/{\rm m_{\rm e} c^2}$. For spherical geometry of external sources $\nu_{\rm ext}' = \Gamma \nu_{\rm ext}$ and for planar geometry $\nu_{\rm ext}' = \Gamma(1-\beta \cos\theta_{\rm m}) \nu_{\rm ext}$, where $\nu_{\rm ext}$ and $\nu_{\rm ext}'$ are the characteristic external photon frequencies in the external and jet co-moving frames, respectively and $\theta_{\rm m}$ is the angle from which contribution to the energy density in the jet co-moving frame is maximal. Radiation energy densities, $u_{\rm ext}'$, of external spherical sources are calculated using the approximate formulas given in Appendix A, and of planar sources  using equations presented by \citet{Sik13} in Appendix A.1. 

\subsection{Radiation}

In the jet co-moving frame radiation produced within an i-th cell is
\beq
\left( \frac {\partial L_{\nu'}'}{\partial \Omega'} \right)_i 
= \int 
{\left( \frac {\partial P_{\nu'}'(\gamma)}{\partial \Omega'} \right)_i} 
N_{\gamma,i} {\rm d}\gamma \, ,
\eeq
where  the single electron radiative power, $\partial P_{\nu'}'(\gamma)/\partial \Omega'$, is superposed from synchrotron, SSC, and ERC components, the latter consisting of radiation from the accretion disk, from broad emission region, and from hot dust. We assume that a conical jet is transversly uniform, so that evolution of the electron energy distribution $N_{\gamma,i}$ is the same along all equally normalised $\theta$-angles within the jet. Therefore, radiation produced at a given distance from the central black hole can be approximated as being  independent of $\theta$. \footnote{A non-zero jet opening angle $\theta_{\rm j}$ implies a spread of Doppler factors for radiation produced by jet elements moving at different angles to the observer's line of sight. It has significant effect on the time profiles of produced flares \citep{Sik01}, yet does not affect time-averaged spectra much.} The total apparent luminosity is then a sum of radiation from $s$ cells
\beq
\nu L_{{\rm app},\nu} =
4 \pi \left(\nu \frac {\partial L_{\nu}}{\partial \Omega} \right) =
4 \pi \frac{{\cal D}^3}{\Gamma} 
\left( \nu' \frac {\partial L_{\nu'}'}{\partial \Omega'} \right) \, ,
\eeq
where
\beq 
\frac {\partial L_{\nu'}'}{\partial \Omega'} = 
\sum_{i=1}^s{
\left( \frac {\partial L_{\nu'}'}{\partial \Omega'} \right)_i } \, ,
\eeq
${\cal D} = [\Gamma(1-\beta\cos{\theta_{\rm obs}})]^{-1}$. 

The synchrotron and SSC luminosities are calculated using the procedure presented in \citet{Mod03}. The ERC luminosities are computed using equations presented in Appendix B. 

\section{Model parameters}

We compute theoretical blazar spectra as a function of the distance from the central black hole $r$, for three different jet magnetization values: $\sigma=1.0$, $0.1$, and $0.01$ and for two geometries of external photon sources. We cover four distance decades, from $10^{16}\,$cm  up to $10^{20}\,$cm. We assume a central black hole mass $M_{\rm BH} = 10^{9} M_{\sun}$, accretion rate $\dot{M} = 3 L_{\rm Edd}/c^2$ and accretion disk radiative efficiency $\eta_{\rm d} \equiv L_{\rm d}/\dot M c^2 = 0.1$ which gives a total accretion disk luminosity $L_{\rm d} \approx 4 \times 10^{46}\, {\rm erg \, s}^{-1}$. With such a value the sublimation radius $r_{\rm sub} \approx 1\,$ pc (see Eq.~\ref{eq:rsub}). Following the results of numerical simulations of magnetically-arrested disks \citep{McK12} and noticing the observed energetics of jets in radio-loud quasars \citep[see][and refs. therein]{SikBeg13} we set the jet production efficiency $\eta_{\rm j} \equiv L_{\rm j}/\dot M c^2 = 1$ which leads to $L_{{\rm j},0} \approx 2 \times 10^{47} \, {\rm erg \, s}^{-1}$. 

The total efficiency of energy dissipation $\eta_{\rm diss}$ must be high because of the high observed $\gamma$-ray luminosities in FSRQs but should not exceed $\sim 0.5$ as a substantial part of the jet energy needs to be transported to radio lobes of FRII radio sources associated with radio-loud quasars.  We set $\eta_{\rm diss} = 0.3$. We use $\Gamma = 15$ \citep{Hov09} and the jet opening angle $\theta_{\rm j} = 1/\Gamma$. Noting 
the indicated by particle-in cell (PIC) simulations of shocks the strong coupling between electrons and protons \citep{Sir11} we assume that the dissipated energy is equally distributed between these particles by setting $\eta_{\rm e} = 0.5$.

The injected electron energy spectrum is assumed to be a broken power law, with spectral indices $p_{1} = -1$ for $\gamma \le \gamma_{\rm b}$ and $p_{2} = 2.5$ for $\gamma > \gamma_{\rm b}$. The choice of a very hard, low-energy injection function  is dictated by the aforementioned energetic coupling between electrons and protons, while a much steeper high-energy portion of the electron injection function is required to reproduce the typical slopes of 
$\gamma$-ray spectra observed by {\it Fermi}/LAT \citep{Ack11}. The break energy $\gamma_{\rm b}$ is calculated using Eqs.~(\ref{eq:deff}), (\ref{eq:gammab}), and~(\ref{eq:gammainj}). 

The detailed parameters used in our modelling are summarised in Tab.~\ref{tab:param}. 

\setcounter{table}{0}
\begin{table*}
\caption{Parameters used in numerical simulations.}
\begin{center}
\begin{tabular}{p{0.5\textwidth}@{}c p{0.5\textwidth}}
\hline
Parameter \hfill & Value \\
\hline
Black hole mass $M_{\rm BH}$ \dotfill & $10^{9} M_{\sun}$ \\ 
Accretion rate $\dot{M}$ \dotfill & $3 L_{\rm Edd}/c^2$ \\ 
Accretion disk radiative efficiency $\eta_{\rm d}$ \dotfill & $0.1$ \\
Jet production efficiency $\eta_{\rm j}$ \dotfill & $1.0$ \\
Energy dissipation efficiency $\eta_{\rm diss}$ \dotfill & $0.3$ \\
Jet Lorentz factor $\Gamma$ \dotfill & $15$ \\
Fraction of energy transferred to electrons $\eta_{\rm e}$ & $0.5$ \\
Jet magnetization $\sigma$ \dotfill & $0.01, 0.1, 1.0$ \\
Pair content $n_{\rm e}/n_{\rm p}$ \dotfill & $1.0$ \\
Electron injection function indices $p_{1}, p_{2}$ \dotfill & $-1.0, 2.5$ \\
Min. and max. injection energies $\gamma_{\rm min}, \gamma_{\rm max}$ \dotfill & $1, 4 \times 10^{4}$ \\
Jet opening angle $\theta_{\rm j}$ \dotfill & $1/\Gamma$ \\
Observing angle $\theta_{\rm obs}$ \dotfill & $1/\Gamma$ \\
BLR photons energy \dotfill & $10\,{\rm eV}$ \\
HDR photons energy \dotfill & $0.06 - 0.6 \, {\rm eV}$ \\
BLR radius \dotfill & $0.1\, r_{\rm sub}$ \\
HDR radius \dotfill & $1.0\, r_{\rm sub}$ \\
BLR covering factor $\xi_{\rm BLR}$ \dotfill  & $0.1$ \\
HDR covering factor $\xi_{\rm HDR}$ \dotfill & $0.3$ \\
\hline
\end{tabular}
\end{center}
\label{tab:param}
\end{table*}

It should be emphasised here that contrary to most studies of blazar spectra, we use the magnetization parameter, $\sigma$, as an input parameter, instead of magnetic field intensity $B'$ or its energy density $u_{\rm B}' = {B'}^2/(8\pi)$. With our input parameters the value of $u_{\rm B}'$ is determined by the following relation
\beq  
u_{\rm B}'= \frac{L_{\rm B}}{\kappa \pi r^2 (\theta_{\rm j} \Gamma)^2 c} =
\frac{\sigma}{1+\sigma} \, 
\frac{(1-\eta_{\rm diss}) L_{{\rm j},0}}{\kappa \pi r^2 (\theta_{\rm j}\Gamma)^2 c},
\label{eq:ub}
 \eeq
where $L_{{\rm j},0} = 0.5 \eta_{\rm j} \dot m L_{\rm Edd}$ (the value of $\kappa$ depends on the ratio of the chaotic to the toroidal magnetic component intensity and is enclosed between $4/3$ and $2$).

\section{Results}

\subsection{Energy densities of external radiation fields}

If spherically isotropized at a distance $r$, the entire disk radiation would have (in a jet co-moving frame) energy density $\simeq L_{\rm d} \Gamma^2/(4\pi r^2 c)$. Hence, this value sets an upper limit for any contribution to $u_{\rm ext}'$ from disk radiation and its fractions $\xi_{\rm{ext}}$ reprocessed in the BLR and HDR. We visualise these contributions in Fig.~\ref{fig:dzeta} as a function of distance using a parameter
\beq
\zeta = \frac{4\pi r^2 {\rm c} u_{\rm ext}'}{L_{\rm d} \Gamma^2} \, .
\label{eq:dzeta}
\eeq
For BLR and HDR $\zeta = g_u \xi_{\rm ext}$, where $g_{\rm u}$ accounts for geometry of a source and was introduced and preliminarily studied by \citet{Sik13} and \citet{Nal14b}. Figure~\ref{fig:dzeta} presents the $\zeta$ parameter for spherical and planar geometries of the BLR and HDR and geometrically thin accretion disk. 

\setcounter{figure}{1}
\begin{figure*}
  \includegraphics{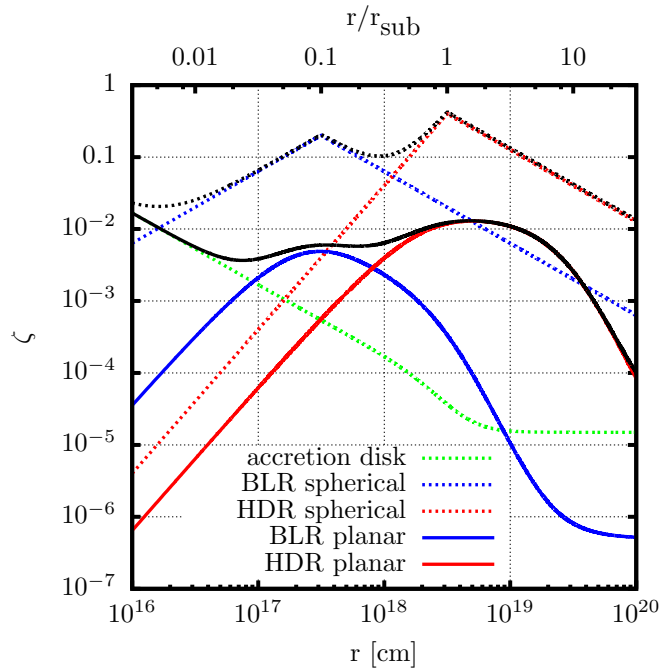}
  \caption{Values of $\zeta$ for spherical and flat geometries of the BLR and HDR and flat accretion disk. The black dotted line is a total $\zeta$ for spherical case and the black solid line presents a total $\zeta$ for planar geometry.}
  \label{fig:dzeta} 
\end{figure*}

Close to the black hole, at $r < 0.01\,$pc, the contribution to $\zeta$ is dominated by the accretion disk and was considered by \citet{Der93} to be the dominant source of seed photons for the ERC process. However, the absence of a bulk Compton feature in the X-ray band indicates that jets at such distances are not yet accelerated enough and, therefore, radiation produced at such distances is not sufficiently Doppler boosted to explain very large luminosities of FSRQs \citep{Sik05, Cel07}. Furthermore, $\gamma$-rays produced too close to the BH would be absorbed  by X-rays from the accretion disk corona \citep[see][and refs. therein]{Ghi12}.

At larger distances the accretion disc contribution to $u_{\rm ext}'$ drops quickly and the BLR starts to dominate. For the spherical model of the BLR its contribution is comparable to that of the accretion disk already at $0.01\,$pc, while for the planar model that distance is about 3 times larger. The maximal BLR contribution to $\zeta$ in both cases is at $r_{\rm BLR} \sim 0.1\,$pc which corresponds with a distance of maximal bolometric BLR luminosity in the calculated spectra. At $r > 0.3\,$pc the contribution to $\zeta$ starts to be dominated by the HDR and reaches a maximum around $1\,$pc, just beyond $r_{\rm sub}$. Then the value of $\zeta$ drops steeply, for the spherical model due to the assumed gradient of the dust radiation luminosity, for the planar model due to assumed presence of outer edge of the HDR (see Appendix A).

It is interesting to note that due to stratification of the BLR and HDR the value of $\zeta_{\rm BLR} + \zeta_{\rm HDR}$ within a distance range $0.03\,$pc - $3\,$pc does not vary strongly with a distance, oscillating around $0.1$ for spherical geometry and around $0.01$ for planar geometry.

\subsection{Bolometric apparent luminosities}

\setcounter{figure}{2}
\begin{figure*}
  \includegraphics[width=0.4\textwidth]{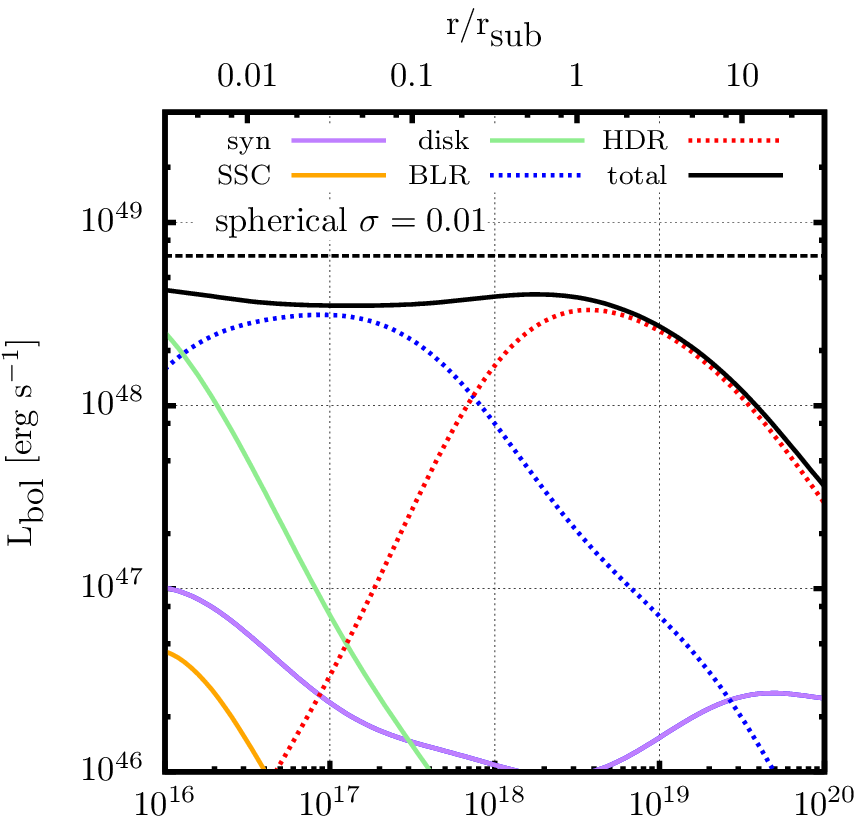}
  \includegraphics[width=0.4\textwidth]{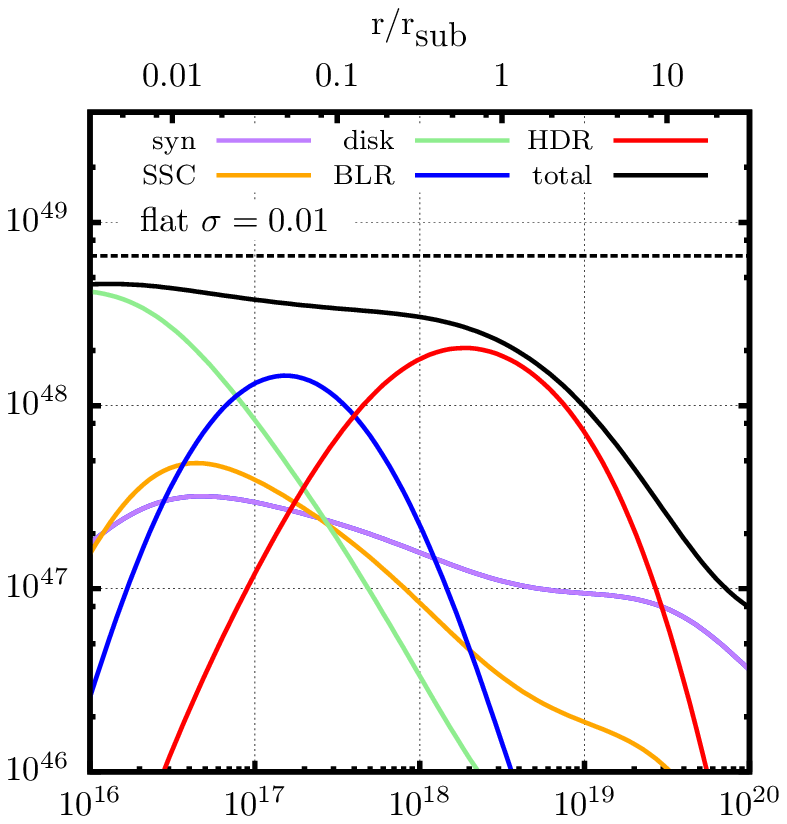} \\
  \includegraphics[width=0.4\textwidth]{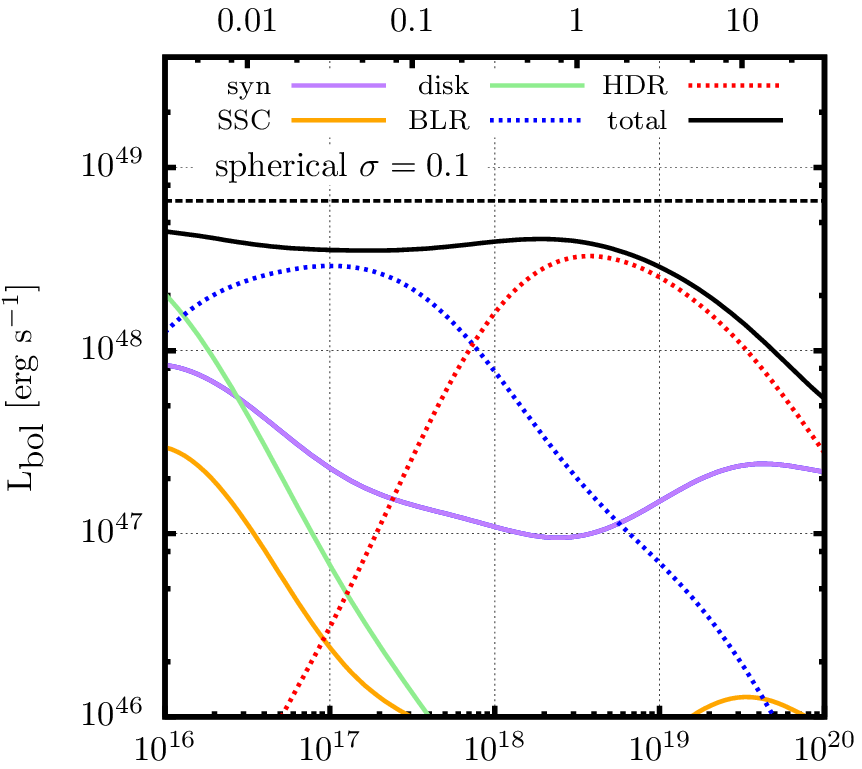}
  \includegraphics[width=0.4\textwidth]{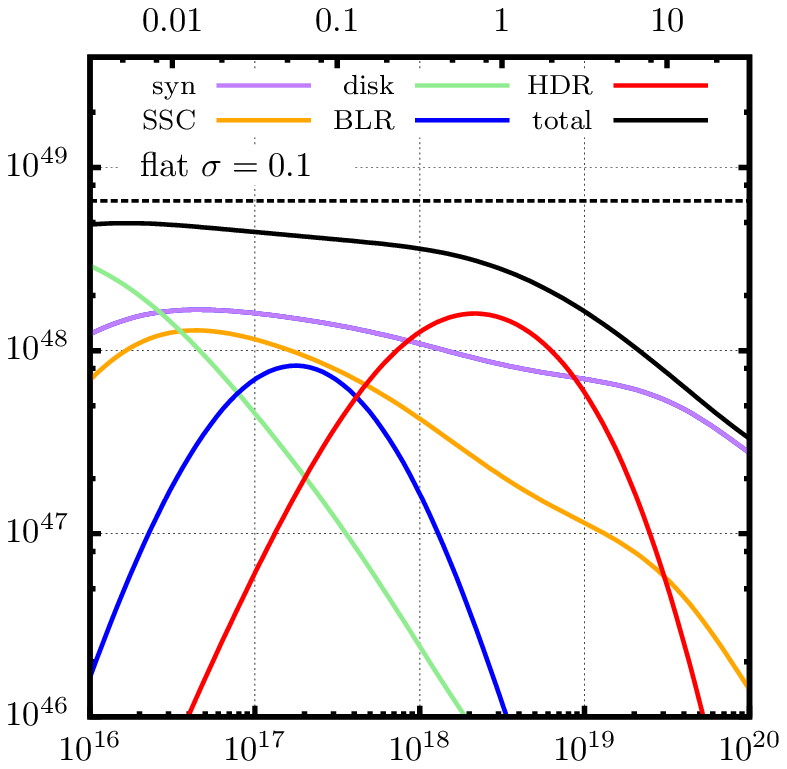} \\
  \includegraphics[width=0.4\textwidth]{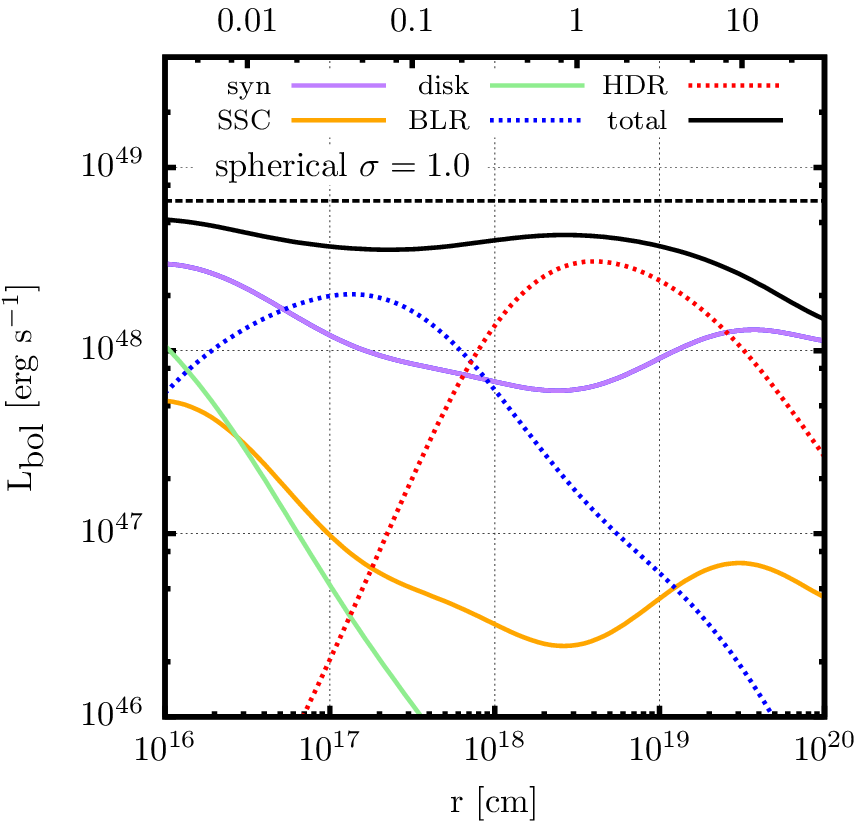}
  \includegraphics[width=0.4\textwidth]{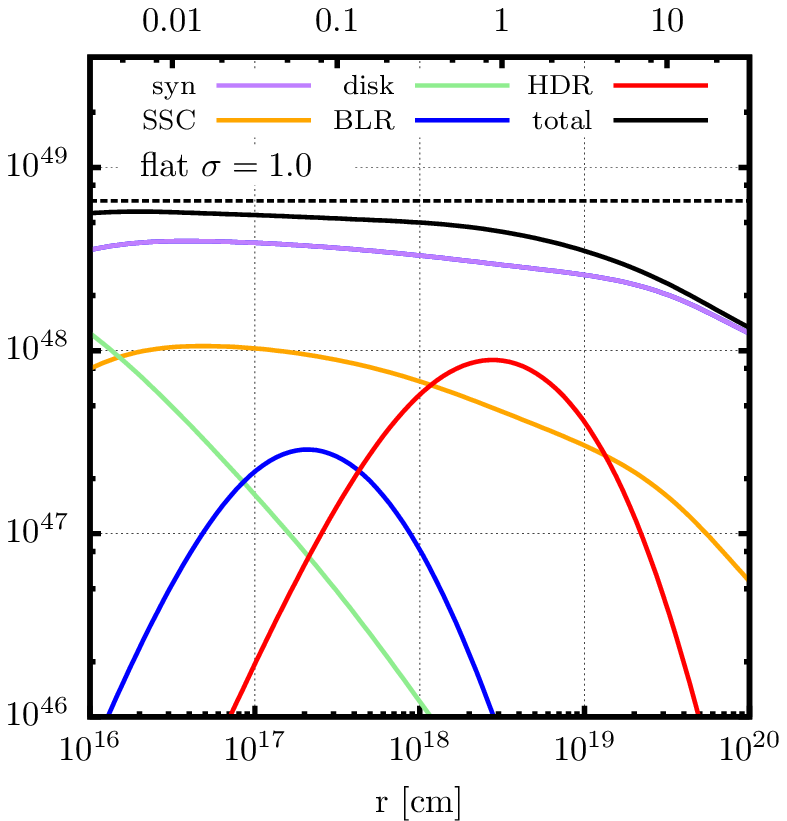} \\
  \caption{Apparent bolometric luminosities calculated for spherical (left column) and flat (right column) geometry of the BLR and HDR and flat accretion disk for $\sigma$ $=0.01$ (1st row), $=0.1$ (2nd row) and $=1.0$ (3rd row). BLR and HDR stand for ERC(BLR) and ERC(HDR), respectively. The black solid line corresponds to total radiative power and the horizontal black dotted line shows total dissipated power transferred to electrons.}
  \label{fig:Lbol} 
\end{figure*}

We calculated apparent bolometric luminosities $L_{\rm bol} = \int L_{\nu} {\rm d} \nu$ for each radiation process separately and their sum \ie~the total radiative power. Figure~\ref{fig:Lbol} presents their dependence on the distance from the central black hole $r$ for three different magnetization parameters and two extreme  geometries. Studying bolometric luminosities is a useful tool to investigate  radiative efficiency and the dominant radiative mechanism where dissipated energy is deposited. For our models the total value of 
dissipated energy channelled to relativistic electrons is $L_{\rm el} \approx 6 \times 10^{48} \,$ erg s$^{-1}$. If the total cooling rate is high enough to effectively cool electrons down to energies $\gamma < \gamma_{\rm b}$, the total radiative efficiency is very high, of the order of $70-90 \%$. It means that the total radiation power is of the order of $10^{48}$ erg s$^{-1}$ as typically observed in FSRQs with $M_{\rm BH} = 10^9 M_{\odot}$. Such a high radiative efficiency is achievable up to distances $\sim 3\,$pc for spherical models and  $\sim 1\,$pc for planar models. At larger distances $\gamma_{\rm cool} > \gamma_{\rm b}$ (where $\gamma_{\rm cool}$ is the electron energy at which the time-scale of radiative energy losses equals the time-scale of adiabatic energy losses) and radiation production efficiency drops quickly. Noting that at such distances $\zeta_{\rm HDR}$ decreases faster than $\zeta_{\rm syn}$, dominance of HDR luminosities drops even faster.

\subsection{Spectral peaks}

\setcounter{figure}{3}
\begin{figure*}
  \includegraphics[width=0.45\textwidth]{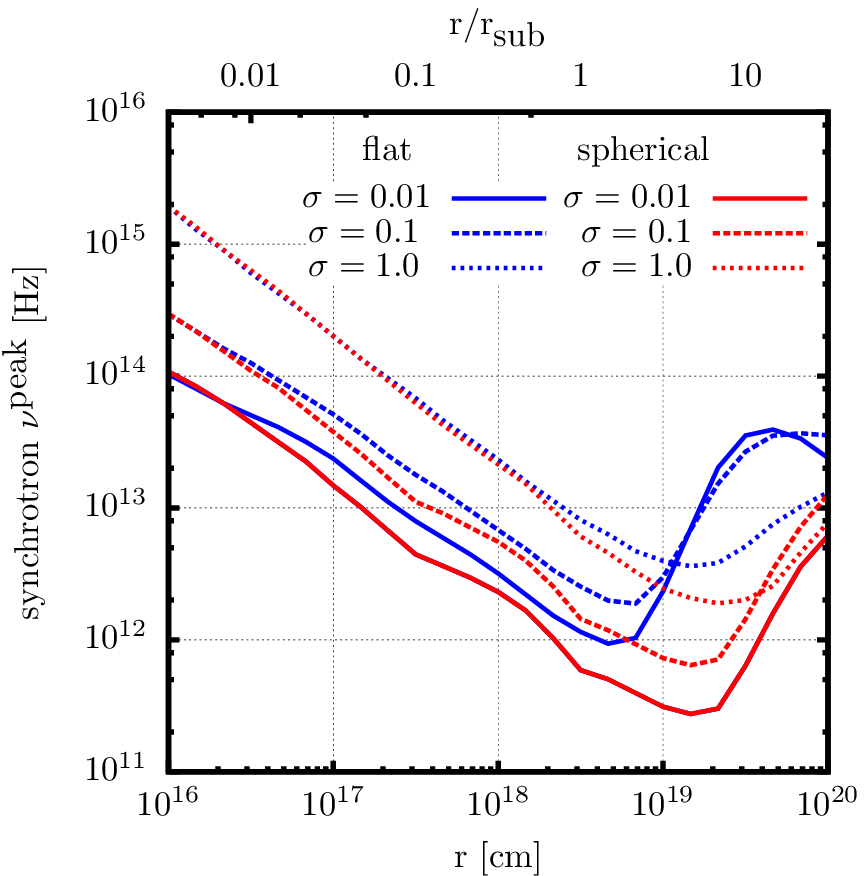}
  \includegraphics[width=0.45\textwidth]{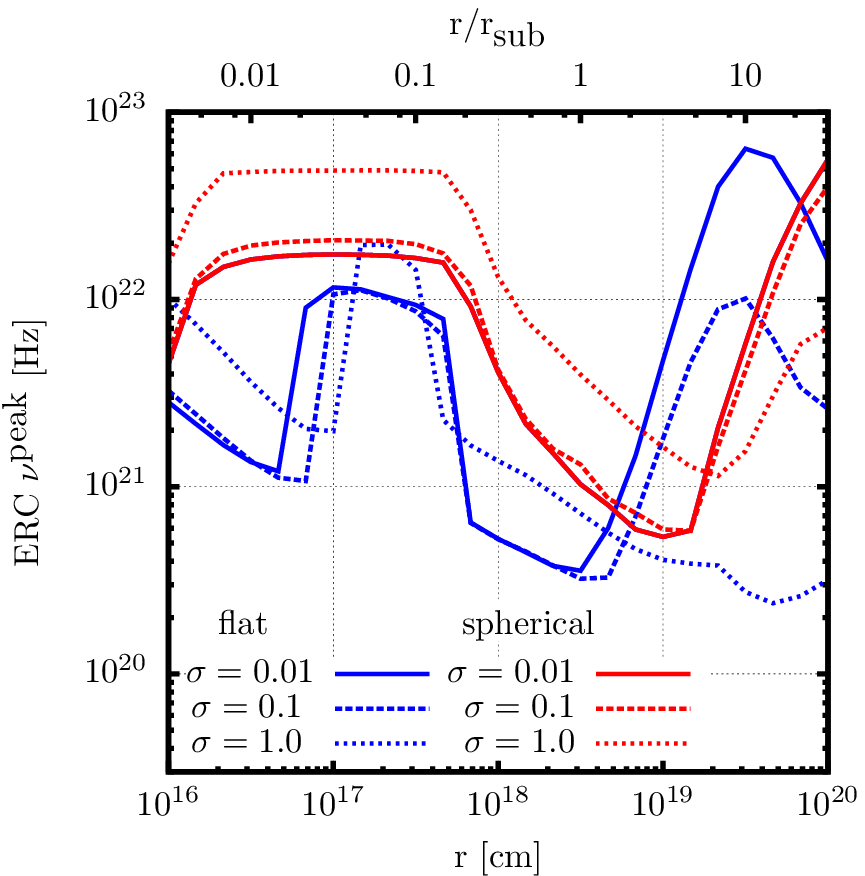}
  \caption{Location of spectral peaks calculated for synchrotron (left panel) and ERC components (right panel) for different values of $\sigma$ and for both planar and spherical geometries of BLR and HDR and flat accretion disk.}
  \label{fig:peaks} 
\end{figure*}

The location of the luminosity peaks of  synchrotron and ERC spectral components in $\nu L_{\nu}$ representation is presented in Fig.~\ref{fig:peaks}. The synchrotron peak is located at $\nu^{peak}_{\rm syn} = 3.7 \times 10^{6} \hat{\gamma}^2 B' \Gamma$, where $B'= \sqrt{8\pi u_{\rm B}'}$, and  $u_{\rm B}'$ is given by Eq.~(\ref{eq:ub}); the ERC peak is at $\nu^{peak}_{\rm ERC} = (16/9) \hat{\gamma}^2 \Gamma^2 \nu_{\rm ext}$, where  
\beq
\hat{\gamma} = \left\{ 
\begin{array}{llll} 
\gamma_{\rm b} & \textrm{for} & \gamma_{\rm cool} < \gamma_{\rm b} & \textrm{(\ie~the fast-cooling regime)}\\
\gamma_{\rm cool} & \textrm{for} & \gamma_{\rm cool}>\gamma_{\rm b} & \textrm{(\ie~the slow-cooling regime)}\\
\end{array}
\right. ,
\eeq
where $\gamma_{\rm b}$ is given by Eqs.~(\ref{eq:deff}), (\ref{eq:gammab}), and~(\ref{eq:gammainj}). As long as electrons with energies $\gamma_{\rm b}$ cool efficiently, the synchrotron peak location drops with distance  
like magnetic field strength \ie~$r^{-1}$, from $10^{14}-10^{15}\,{\rm Hz}$ at $10^{16}\,$cm to $10^{12}-10^{13}\,$Hz at $\sim 1\,$pc. At larger distances the synchrotron peaks are produced by electrons with energies $\gamma_{\rm cool}$ and, therefore, their location shifts very fast to higher frequencies, until $\gamma_{\rm cool} = \gamma_{\rm max}$ at $r \sim 10\,$pc.

The location of ERC luminosity peaks depends on a dominant source of seed photons at a particular distance $r$ and their characteristic energy, h$\nu_{\rm ext}$. The largest values are observed at distances of ERC(BLR) domination and $\gamma$-ray peaks result from Compton up-scattering of photons with energies  $\sim 10\,$eV of most luminous spectral lines. At $r < 0.03\,$pc for planar BLR geometry we can see the effect of the ERC(disk) domination. Much lower -- but increasing with decreasing distance -- ERC peak location is determined by the temperature of these portions of the accretion disk which contribute the most to $u_{\rm ext}'$ at a given distance
along a jet. At distances between $0.3$ and $3\,$pc the $\gamma$-ray luminosity peak is set by ERC(HDR), with energies of the seed photons h$\nu_{\rm ext} \sim 0.3\,$eV. At $r>3\,$pc we see the same effect as in the case of the synchrotron peak \ie~the luminosity peak is produced by electrons with energies $\gamma_{\rm cool}$ and therefore its location increases with distance very fast up to the distance where $\gamma_{\rm cool}= \gamma_{\rm max}$ and then sharply drops. For all models taken into consideration $10^{20}\,{\rm Hz} < \nu^{\rm peak}_{\rm ERC} < 10^{23}\,$Hz.

\subsection{Compton dominance}

\setcounter{figure}{4}
\begin{figure*}
  \includegraphics[]{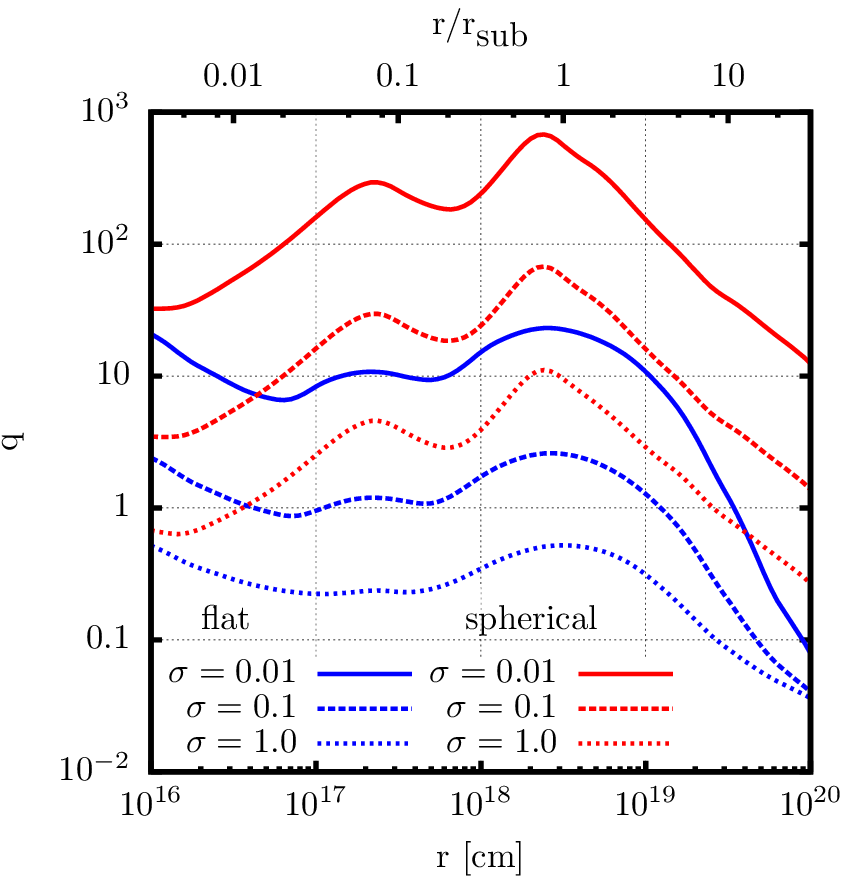}
  \caption{Compton dominance parameter $q$ for different external source geometries and values of $\sigma$.}
  \label{fig:q} 
\end{figure*}

The Compton dominance parameter $q$ is defined as the ratio of the ERC-to-synchrotron peak luminosities, $(\nu L_{\nu})_{\rm ERC}^{\rm peak}/ (\nu L_{\nu})_{\rm syn}^{\rm peak}$. Its dependence on a distance from the BH
for different parameters $\sigma$ and different external radiation source geometries is shown
in Fig.~\ref{fig:q}. One can easily notice that the shape of $q(r)$ is very  similar to the shape of $\zeta(r)$ (see Fig.~\ref{fig:dzeta}). This can be explained by using approximate scalings of peak luminosities $(\nu L_{\nu})_{\rm ERC}^{\rm peak} \propto u_{\rm ext}'$ and $(\nu L_{\nu})_{\rm syn}^{\rm peak} \propto u_{\rm B}'$. Using Eqs.~(\ref{eq:ub}) and~(\ref{eq:dzeta}) one can approximate $q$ by the formula
\beq
q \sim \frac{u'_{\rm ext}}{u'_{\rm B}} = 
\frac{(1+\sigma)\Gamma^2 (\theta_{\rm j} \Gamma)^2 \zeta \eta_{\rm d}}{4\sigma \eta_{\rm j}}
\, .
\label{eq:q}
\eeq
Hence, for the fixed values of $\Gamma$, $\theta_{\rm j}$, $\eta_{\rm j}$, and $\eta_{\rm d}$ the approximate dependence of $q$ on distance is the same as of $\zeta$, while its normalisation depends on $\sigma$. We can see in Fig.~\ref{fig:q} that for planar geometries values of $q>4$, typical for FSRQs, are achievable only for $\sigma < 0.03$. In case of spherical models and $r > 0.01\,$pc, large values of Compton dominance require $\sigma <0.3$. All the discussed spectral features for a whole range of studied distances and parameters are presented in Fig.~\ref{fig:sed} where examples of computed broad-band spectra are presented.

\setcounter{figure}{5}
\begin{figure*}
  \includegraphics[width=0.45\textwidth]{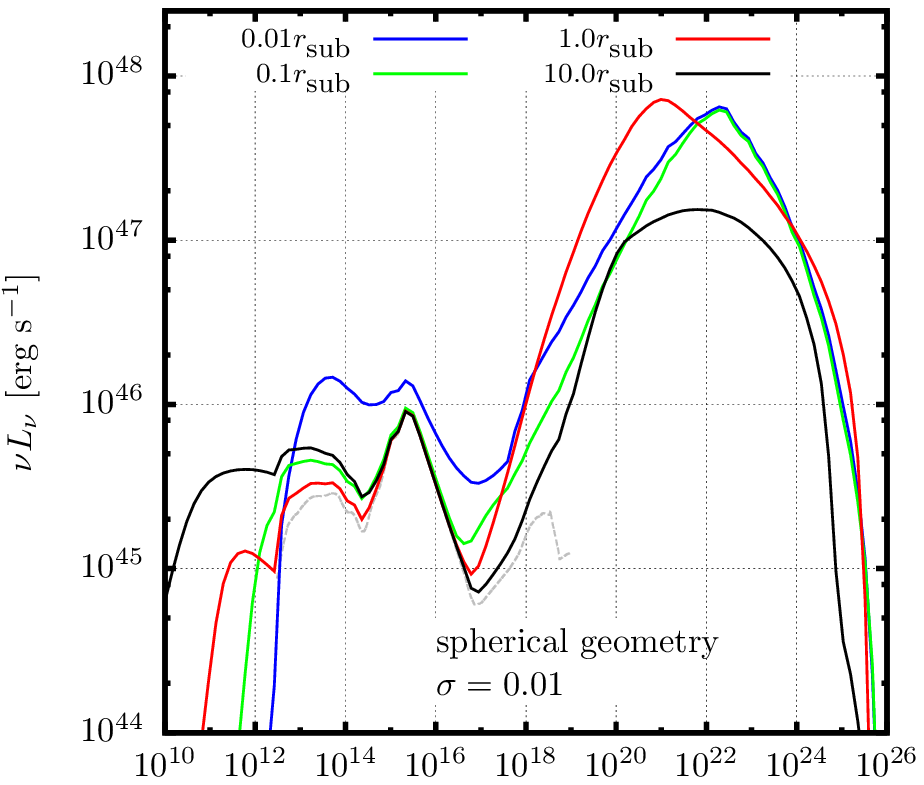}
  \includegraphics[width=0.45\textwidth]{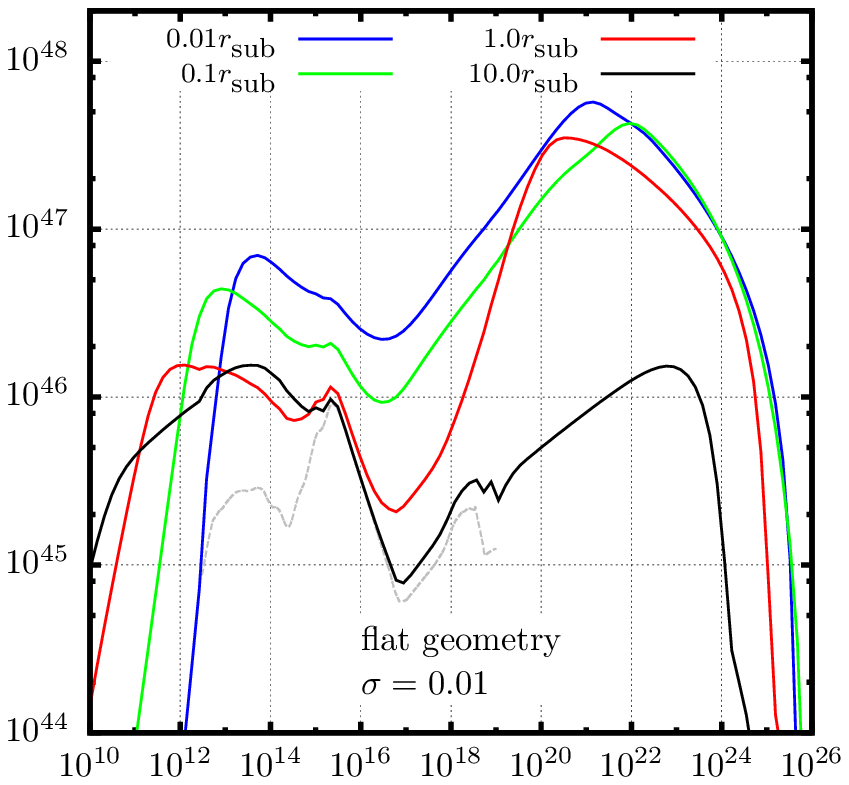} \\
  \includegraphics[width=0.45\textwidth]{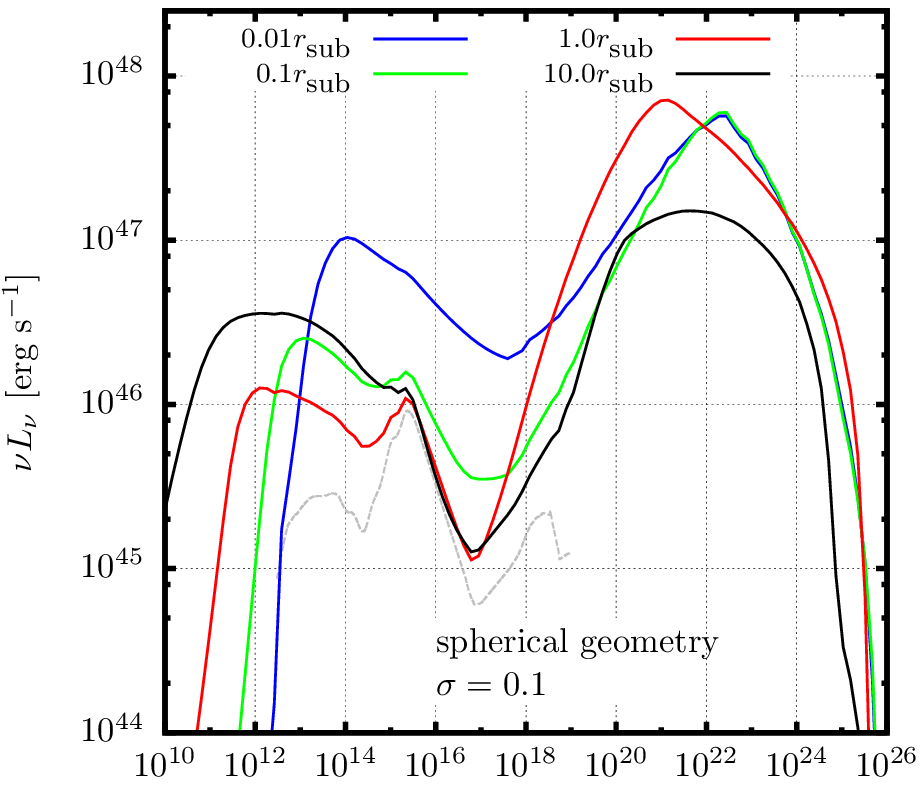} 
  \includegraphics[width=0.45\textwidth]{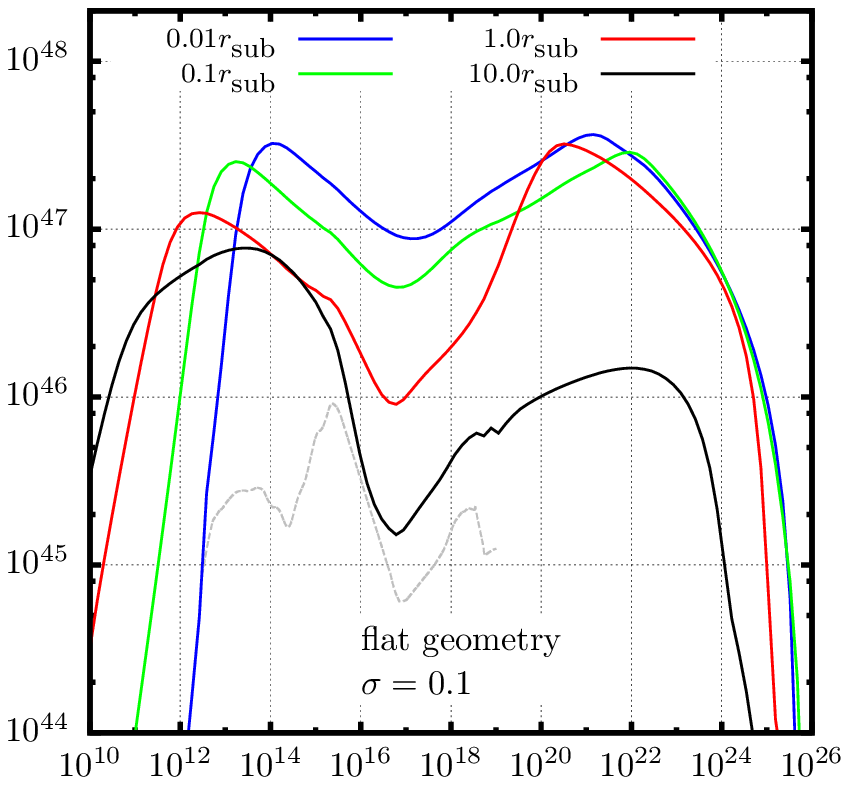} \\
  \includegraphics[width=0.45\textwidth]{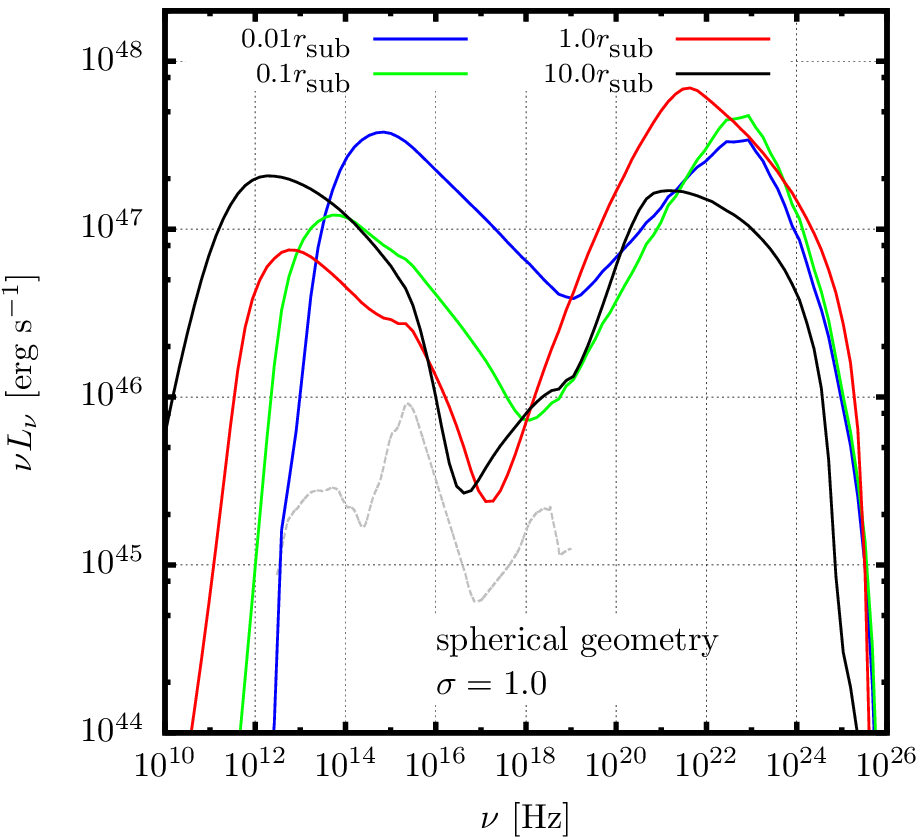}
  \includegraphics[width=0.45\textwidth]{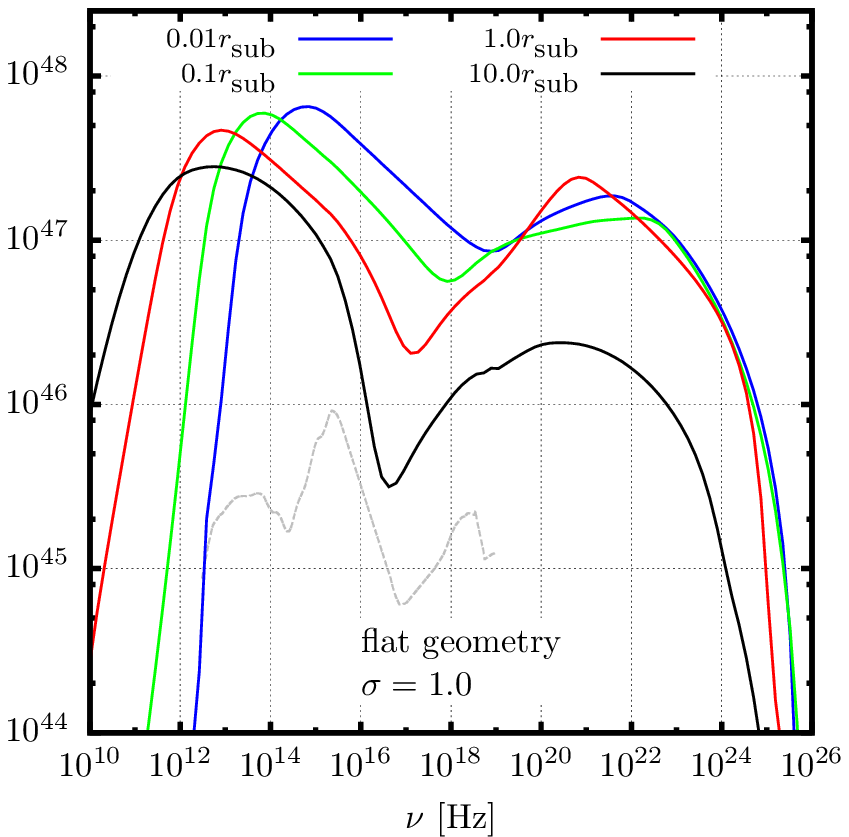} \\
  \caption{Spectral energy distributions calculated for spherical (left column) and flat (right column) geometry of the BLR and HDR and flat accretion disk for $\sigma$ $=0.01$ (1st row), $=0.1$ (2nd row) and $=1.0$ (3rd row). Solid lines corresponds to the sum of all radiation components and the grey dotted line shows radio-loud quasar radiation template \citep{Elv94}.}
  \label{fig:sed} 
\end{figure*}

\section{Discussion}

The most diverse opinions about the nature of AGN relativistic jets concern their magnetization, $\sigma$. Being initially dominated by the Poynting flux, relativistic jets are thought to be converted at some distance to matter-dominated flows \citep[see][and refs. therein]{Sik05}. The conversion process can initially proceed pretty efficiently even if the jet is stable and in a steady-state \citep{Kom09, Tch09, Lyu10a}. However, after $\sigma$ drops to unity the conversion process becomes inefficient and magnetization decreases much slower with the distance unless being supported by such processes as MHD instabilities, randomisation of magnetic fields \citep{Hei00}, reconnection of magnetic fields \citep{Dre02, Lyu10b}, and/or impulsive modulation of jet production \citep{Lyu10, Gra11}. Little is known about the feasibility and efficiency of these processes in context of AGN and the only chance to verify evolution of magnetization in relativistic jets is by investigating their observational properties over different spatial scales. Close to the jet base such studies can only be performed by analysing the broad-band spectra in blazars.

Results obtained using the ERC models for $\gamma$-ray production in FSRQs indicate that jets in the blazar zones of these objects are dominated by an energy flux of cold protons (see \citealt{Ghi10}; \citealt{Ghi14}, and refs. therein). These results have been obtained by recovering the jet parameters and the location of the blazar-zone from the fits of the observed spectra. However, noticing the very poor coverage of the blazar spectra in the far IR and in the $10\,$keV - $100\,$MeV band where the synchrotron and ERC luminosity peaks in FSRQs are usually located, the quality of such fits is very limited. 



In this paper we present the results of modelling blazar spectra with parameters typical for radio-loud quasars (see Table 1) and assuming (suggested by PIC simulations) strong coupling between electrons and protons heated in shocks \citep{Sir11}. The blazar spectra were computed as a function of distance for three different values of $\sigma$, $1$, $0.1$ and $0.01$. Since the geometry of the BLR and HDR is not well determined we took into consideration both flat and spherical geometry with radial stratification. Any kind of "real" geometry of the BLR and HDR is thought to be in between those two extreme cases. As a source of seed photons for the ERC process, we also included the accretion disk.

We would like to stress that the presented model and calculations do not use any 'best-fit' values. All used parameters are either observationally determined or predicted by other models and we do not use any model fine-tuning. In fact, Eq.~\ref{eq:q} shows that the only free parameter on which magnetisation is dependent is $\zeta$ (calculated numerically) and $\theta_{\rm j} \Gamma$. Since causality conditions imply $\theta_{\rm j} \Gamma \le 1$ \citep{Cla13}, while constraints imposed by the synchrotron-self-Compton process \citep{Nal14a} and observations of radio cores \citep{Push09} indicate that $\theta_{\rm j} \Gamma \ge 0.1$, in the presented calculations we adopted a value of $\theta_{\rm j} \Gamma = 1$ and in the discussion below we show the approximate dependence of magnetisation constraints on $\theta_{\rm j} \Gamma$ using Eq.~\ref{eq:q}. Other parameters mentioned in Table 1 have no significant impact on these constraints.

The computed models allowed us to calculate constraints on the jet magnetization as imposed by the spectral features produced at distances within the range $10^{16}-10^{20}\,$cm.
Typical for FSRQs, the values of Compton dominance, $4 < q < 30$, are found to be reproducible for $\sigma \sim 0.1(\theta_{\rm j} \Gamma)^2$ in case of the spherical geometries of the BLR and HDR and for $\sigma \sim 0.01(\theta_{\rm j} \Gamma)^2$ in case of the planar geometries (see Fig.~\ref{fig:q} and Eq.~\ref{eq:q}). Noting that the real geometries of the BLR and HDR are intermediate between the spherical and planar, one may conclude that typical values of the Compton dominance imply $\sigma \sim 0.03 (\theta_{\rm j} \Gamma)^2$. This result indicates that the  conversion of the initially Poynting flux dominated jet to the matter  dominated jet takes place in the region located closer to the BH than the blazar zone.

Our studies also provide interesting constraints on other model parameters. This particularly concerns maximal and minimal distances of the blazar zone location and pair contents. Due to the fast drop of radiation efficiency at large distances, the blazar zone should not be located farther than $\sim 3\,$pc (see \S{4.2} and Figs.~\ref{fig:Lbol} and~\ref{fig:sed}); due to an increase of the synchrotron peak frequency with decreasing distance the typical observed value $\nu_{\rm syn,peak} \sim 10^{13}\,$Hz \citep[see][Fig. 6]{Ack11} is achievable at $r> 10^{17}\,$cm, unless $n_{\rm e}/n_{\rm p} \gg 1$ (see the left panel in the Fig.~\ref{fig:peaks} and Eq.~\ref{eq:gammainj}). Contrary, in the ERC(HDR) dominance region, the pair content is required to be negligible in order to keep location of the ERC spectral peaks within the observationally acceptable range (see the right panel in the Fig.~\ref{fig:peaks} and \citealt[Fig.~18]{Gio12}).

\section{Conclusions}

Our main results can be stressed as follows:

\begin{description}
\item - large $\nu_{\rm syn}^{\rm peak}$ at distances $r < 0.03\,$ pc (\S{3.3} and Fig.~\ref{fig:peaks}) and low radiative efficiencies at $r > 3\,$ pc (\S{3.2} and Fig.~\ref{fig:Lbol}) seem to favour the location of the blazar zone in FSRQs within a distance range $0.03 - 3\,$ pc; this range is similar to the one obtained by \citet{Nal14a} by using variability and compactness constraints;  

\item - typical values of FSRQ Compton dominance parameter, $q \sim 10$, can be recovered within a distance range $0.03 - 3\,$ pc for $\sigma \sim 0.1(\theta_{\rm j} \Gamma)^2$ in case of spherical BLR and HDR and $\sigma \sim 0.01(\theta_{\rm j} \Gamma)^2$ for their planar geometries (see Fig.~\ref{fig:q} and Eq.~\ref{eq:q});

\item - due to the value of $\gamma_{\rm b}$ being fixed by the fixed dissipation efficiency $\eta_{\rm diss}$ (see \S{2.2}) and $\nu_{\rm BLR}/\nu_{\rm HDR} \sim 30$, the spectral peak of ERC(HDR) is located at $\sim 30$ times lower energy than the spectral peak of ERC(BLR) (see \S{4.3} and Fig.~\ref{fig:peaks}); noting that $\nu_{\rm HDR}^{\rm peak} \sim (n_{\rm p}/n_{\rm e})^2\,$MeV, the ERC(HDR) models with significant pair content are rather excluded;

\item - noting that the real geometries of the BLR and HDR are suggested by observations to be significantly flattened (see \S{2}), one may conclude that typical values of the Compton dominance imply $\sigma \ll 1$, and therefore, that the conversion of initially Poynting flux dominated jet to the matter dominated jet takes place in the region located closer to the BH than the blazar zone.
\end{description}

Obtained constraints on the jet magnetization can be at least quantitatively relaxed, if noting the possibility that jets in blazars are magnetically very inhomogeneous and that most blazar emission takes place in weaker magnetized sites being associated with reconnection layers and/or the jet spine region. Such a possibility was investigated by \citet{Nal14b} following claims that value of $\sigma$ in radio cores is of the order of unity \citep{Zam14}, however found by \citet{Zdz14} to be overestimated.

\section*{Acknowledgements}
 
We thank the referee for critical comments which helped to improve the paper. We acknowledge financial support by the Polish NCN grants DEC-2011/01/B/ST9/04845, DEC-2012/07/N/ST9/04242, and DEC-2013/08/A/ST9/00795, the NSF grant AST-0907872, the NASA ATP grant NNX09AG02G.

\appendix

\section{Radiation energy densities of external sources}

\begin{subsection}{Spherical geometry of external sources}

In the case of spherical geometry we assume approximate stratification of BLR
emission by two functions, $\partial L_{\rm BLR}/\partial {\rm ln} r \propto r$
for $r < r_{\rm BLR}$, and $\partial L_{\rm BLR}/\partial {\rm ln} r \propto 1/r$ 
for
$r>r_{\rm BLR}$, where $r_{\rm BLR}$ is the distance at which the maximum BLR luminosity
is produced. For such a stratification energy density of BLR radiation field
on the jet axis in the jet co-moving frame is  

\beq
u_{\rm BLR}' = \frac{\xi_{\rm{BLR}} L_{\rm{d}}}{6 \pi {\rm c} r_{\rm BLR}^2} \times 
\left\{ 
\begin{array}{lcl}
\left( \frac{r}{r_{\rm{BLR}}} \right)^{-1} & \rm{for} & r \le r_{\rm{BLR}}\\
\left( \frac{r}{r_{\rm{BLR}}} \right)^{-3} & \rm{for} & r > r_{\rm{BLR}}
\end{array}
\right. \, ,
\eeq
where $\xi_{\rm BLR} \equiv L_{\rm BLR}/L_{\rm d}$. 

For the HDR we assume that $\partial L_{\rm HDR}/\partial {\rm ln}r = 0$ for 
$r<r_{\rm sub}$
and $\partial L_{\rm HDR}/\partial {\rm ln} r \propto 1/r$ for  $r>r_{\rm sub}$.
The jet co-moving energy density of such radiation field is 

\beq
u_{\rm HDR}' = \frac{\xi_{\rm{HDR}} L_{\rm{d}}}{3 \pi  {\rm c} r_{\rm sub}^2} \times
\left\{ 
\begin{array}{lcl}
1 & \rm{for} & r \le r_{\rm{sub}}    \\
\left( \frac{r}{r_{\rm{sub}}} \right)^{-3} & \rm{for} & r > r_{\rm{sub}}
\end{array}
\right. \, ,
\eeq
where $\xi_{\rm HDR} \equiv L_{\rm HDR}/L_{\rm d}$ and  
\beq
r_{\rm sub} = 1.6 \times 10^{-5} L_{\rm d}^{1/2}
\label{eq:rsub}
\eeq
is the graphite sublimation radius (Sikora \ea~2013).
\end{subsection}

\begin{subsection}{Planar geometry of external sources}

The energy density of radiation field on the jet axis in the jet co-moving frame is 

\begin{multline}
u_{\rm ext}' = \frac{1}{c} \int{I_{\rm ext}' d\Omega_{\rm ext}'} = \frac{1}{c} \int{\frac{I_{\rm ext}}{{\cal D}_{\rm ext}^2} d\Omega_{\rm ext}} \\
= \frac{\Gamma^2}{4\pi c} \int_{R_1}^{R_2} { \frac{(1-\beta\cos\theta_{\rm ext})^2 f_{\rm d}(\theta_{\rm ext})}{r^2 + R^2} \, \frac{\partial L_{\rm ext}}{\partial R}\,  \rm{d} R} \label{uext} \, ,
\end{multline}
where 
\beq {\cal D}_{\rm ext} \equiv \nu_{\rm ext}/\nu_{\rm ext}' =
[\Gamma(1-\beta \cos \theta_{\rm ext})]^{-1} \, , 
\eeq
\beq \rm{d} \theta_{\rm ext} = \frac {\cos \theta_{\rm ext} \rm{d} R}
{\sqrt{r^2+R^2}} = \frac {r {\rm d} R} {r^2 + R^2} \, , \eeq
\beq {\rm d} \Omega_{\rm ext} = \sin\theta_{\rm ext} {\rm d}\theta_{\rm ext} {\rm d}\phi_{\rm ext}
= \frac{r R \, \rm{d}R \, \rm{d}\phi_{\rm ext}}
{(r^2 + R^2)^{3/2}} \, . \eeq
For the optically thick sources (the case of the accretion disk) $f_{\rm d}(\theta_{\rm ext}) = 2 \cos{\theta_{\rm ext}}$, for optically thin sources (assumed to be the case of the BLR and HDR) $f_{\rm d}=1$. Accretion disk is assumed to extend from $R_1 = R_{\rm g}$ (where $R_{\rm g}$ is the gravitational radius) to $R_2 = r_{\rm sub}$, BLR -- from $R_1 = 0.1 r_{\rm sub}$ to $R_2=r_{\rm sub}$, and HDR -- from $R_1 = r_{\rm sub}$ to $R_2= 10 r_{\rm sub}$ \citep{Sik13}.
\end{subsection}

\section{ERC luminosities}

\begin{subsection}{Spherical geometry of external sources}

The ERC luminosity for spherical external radiation field is (see Eq.~15 in \citealt{Mod05})
\beq
\nu' \frac {\partial L_{\nu'}'} {\partial \Omega'} =
{3 {\rm c} \sigma_T \over 16\pi} u_{\rm{\rm ext}}' \left( \nu' \over \nu_{\rm ext} \right)^2
\int {\frac {N_{\gamma}}{\gamma^2} \, f_{\rm sc} \, {\rm d} \gamma} \, ,
\eeq
where $u_{\rm ext}'$ for BLR and HDR are given in our Appendix A, while function $f_{\rm sc}$ is specified by Eq.~(A3) in \citet{Mod05}.
\end{subsection}

\begin{subsection}{Planar geometry of external sources}

We adopt the formalism presented in \citet{Der02} and \citet{Der09} to calculate ERC radiation for seed photons coming from the flat accretion disk. This work extends it to include seed photons from planar BLR and HDR.

The ERC luminosity produced at a distance $r$ by Comptonization of radiation reaching a jet from different directions is 
\beq
\nu' \frac {\partial L_{\nu'}'} {\partial \Omega'} 
={3\sigma_T \over 16\pi} {\nu'}^2 \int_{4\pi}
 \int
\frac {I_{\nu_{\rm ext}}} {{\nu_{\rm ext}}^2} \, \left[ \int
\frac {N_{\gamma}}{\gamma^2} \,f_{\rm sc} {\rm d}\gamma 
\right] {\rm d}\nu_{\rm ext} {\rm d}\Omega_{\rm ext}  \, , \label{LERC2}  
\eeq
where ${\rm d}\Omega_{\rm ext} = {\rm d} \cos\theta_{\rm ext} {\rm d} \phi_{\rm ext}$, $\theta_{\rm ext}$ and $\phi_{\rm ext}$ are the polar and azimuthal angles of the rays approaching a jet at a given distance $r$ determined in the external (BH) frame. This equation comes from generalisation of Eq.~(15) in \citet{Mod05} to include external radiation coming in a jet co-moving frame from all directions and using the Lorentz invariant $I_{\nu_{\rm ext}} {\rm d}\nu_{\rm ext} {\rm d}\Omega_{\rm ext} /\nu_{\rm ext}^2$. For the planar external sources with  ${\rm d}R$-rings approximated to radiate mono-energetically but with $\nu_{\rm ext}$ being in general dependent on $R$ above expression for the ERC luminosities takes the form
%
\begin{multline}
\nu' \frac {\partial L_{\nu'}'} {\partial \Omega'} = {3\sigma_T \over 8} {\nu'}^2 \int_{R_1}^{R_2} \frac{I_{\rm ext}}{(\nu_{\rm ext})^2} \, \frac{rR}{(r^2+R^2)^{3/2}} \\ 
\times \left[\int_{\gamma_{\rm m}}^{\gamma_{\rm max}} \frac {N_{\gamma}}{\gamma^2}\,f_{\rm sc} {\rm d}\gamma\right] \, {\rm d}R  \, ,  \label{LERC4}  
\end{multline}
where 
\beq I_{\rm ext}  = 
\frac {f_{\rm d}(\theta_{\rm ext})} {8\pi^2 R \cos\theta_{\rm ext}} 
\frac{\partial L_{\rm ext}}{\partial R} \, 
\, , \label{I-planar2} \eeq
and luminosities $\partial L_{\rm ext}/\partial R$ are specified by \citet{Sik13} in Appendix A.2.
\end{subsection}

\end{document}